# On the Testing of complete causal mediation and its applications

Yichin Tsai[a], Wan-Tzu Chang[a], Jia Jyun Sie[b], Cathy SJ Fann[b]*, Iebin Lian[a]*

[a] Institute of Statistics and Information Science, National Changhua University of Education, Taiwan,
[b] Institute of Biomedical Sciences, Academia Sinica, Taipei, Taiwan

*correspondent email: csjfann@ibms.sinica.edu.tw; maiblian@cc.ncue.edu.tw.

ORCID (Iebin Lian) 0000-0003-3992-6049

**Statements and Declarations**

The authors declare no conflict of interest related to the manuscript.

**Funding:**

The work was partially supported by NSTC 113-2314-B-018-001 from the National Science and Technology Council of Taiwan. The funders had no role in study

**Author contributions:**

Conceptualization: [Iebin Lian]; Methodology: [Cathy SJ Fann]; Formal analysis and investigation: [Yichin Tsai, Jia Jyun Sie]; Writing - original draft preparation: [Iebin Lian, Yichin Tsai]; Writing - review and editing: [Jia Jyun Sie]; Funding acquisition: [Iebin Lian, Cathy SJ Fann]; Resources: [Cathy SJ Fann]; Supervision: [Yichin Tsai].

ABSTRACT

The Complete Mediation Test (CMT) serves as a specialized approach of mediation analysis to assess whether an independent variable A, influences an outcome variable Y exclusively through a mediator M, without any direct effect. An application of CMT lies in Mendelian Randomization (MR) studies, where it can be used to investigate non-pleiotropy—that is, to test whether genetic variants impact a disease outcome solely through their effect on a target exposure variable. Traditionally, CMT has relied on two significance-based criteria and a proportion-based criterion with a heuristic threshold that has not been rigorously evaluated.

In this paper, we explored the theoretical properties of conventional CMT, and proposed using standardized absolute proportion of mediation (SAPM) as a criterion for CMT. We, systematically assess the performance of various CMT criteria via simulation, and demonstrate their practical utility in the context of MR studies. Our results indicate that the offers the best performance. We also propose using different optimal thresholds depending on whether the mediator and outcome are continuous or binary. The SAPM with proper thresholds ensures that the indirect pathway meaningfully accounts for the effect of the exposure on the outcome, thereby strengthening the case for complete mediation.

## 1. Introduction

Mediation relationships, or indirect effects, are widely used to elucidate the mechanisms through which effects occur in prevention and intervention research (Fairchild & MacKinnon, 2009). For instance, in studies of vaccine efficacy (Y), the effect of vaccination (A) can be decomposed into a direct effect—such as the T cell-mediated cellular response—and an indirect effect mediated by antibodies (M) from the humoral response (Lin et al., 2023). Mediation analysis methods have been extended from continuous to dichotomous outcomes to accommodate causal mediation frameworks (Imai et al., 2010; VanderWeele & Vansteelandt, 2010). The basic structure of such analyses is illustrated in Fig 1., and the mediation test typically involves the following steps:

(1) Demonstrate that A is associated with Y (via regression),
(2) Show that A is associated with M (regress M on A), and
(3) Show that M is associated with Y, controlling for A (regress Y on A and M).

## 2. Continuous mediator and response

When both M and Y are continuous, the results from causal mediation analysis closely align with those from traditional mediation approaches. For simplicity, we adopt the notation from the traditional framework, in which the mediation (indirect) effect can be derived using linear regression and expressed as $\alpha \cdot \beta = \tau - \tau'$.

To assess whether M partially mediates the causal pathway from X to Y, one tests the hypothesis: $H_0$: 'no indirect effect' or, equivalently, $H_0: \alpha \cdot \beta = 0$ vs. $H_1: \alpha \cdot \beta \neq 0$. Rejection of $H_0$ provides evidence that M mediates the effect of X on Y.

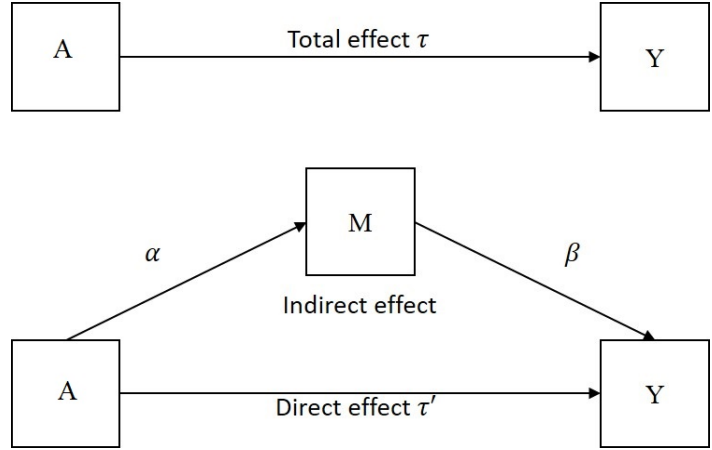

Fig 1. Causal diagram among independent variable A, mediator M and outcome Y.

In contrast to the test for partial mediation described above, complete mediation implies that the effect of A on Y operates entirely through the mediator M (Judd and Kenny, 2010)). Traditionally, complete mediation is supported when the indirect effect is statistically significant, while the direct effect is not (Baron & Kenny, 1986). Formally, the composite hypothesis:

$H_0: \alpha\beta = 0$ or $\tau' \neq 0$ (not complete mediation) vs $H_1: \alpha\beta \neq 0$ and $\tau' = 0$ (complete mediation) can be decomposed into two sub-hypotheses:

$H_{0(i)}: \alpha\beta = 0$ vs $H_{1(i)}: \alpha\beta \neq 0$ (no indirect effect vs with indirect effect)

$H_{0(ii)}: \tau' = 0$ vs $H_{1(ii)}: \tau' \neq 0$. (no direct effect vs with direct effect)

The conventional criteria applied are:

(i) Reject $H_{0(i)}: \alpha\beta = 0$ in favor of $H_{1(i)}: \alpha\beta \neq 0$ by $Z_{IE} = \frac{|\hat{\alpha}\cdot\hat{\beta}|}{se(\hat{\alpha}\cdot\hat{\beta})} > z_{0.025}$, and

(ii) Fail to reject $H_{0(ii)}: \tau' = 0$ in favor of $H_{1(ii)}: \tau' \neq 0$ by $Z_{DE} = \frac{|\hat{\tau'}|}{se(\hat{\tau'})} < z_{0.025}$.

We denoted the complete mediation test composed by criteria (i) and (ii) as *CMT1*. A drawback of this test is the potential for an inflated Type I error rate. Specifically, the actual Type I error rate is probability of rejecting $H_{0(i)}$ (i.e., detecting an indirect effect) while failing to reject $H_{0(ii)}$, given that either $H_{1(i)}$ or $H_{0(ii)}$ is true. It can be expressed as: $P(Z_{IE} > z_{0.025} \cap Z_{DE} < z_{0.025} | \alpha\beta = 0 \cup \tau' \neq 0)$.

The following session explore the theoretical properties such as power and Type I error rate of CMT1.

## 3. Theoretical properties of conventional CMT

Let us denote the standardized direct and indirect effects be $Z_{DE} = Z_1$, $Z_{IE} = Z_2$. Assuming $(Z_1, Z_2) \sim BVN(0, \mu_2, 1, 1, \rho)$, then complete mediation (CM) holds if $\mu_2 \neq 0$ and $\mu_1 = 0$. For hypothesis $H_0$: non−CM vs $H_1$: CM, the criteria for CMT1 is to reject $H_0$ if $|Z_2| > z_{a/2}$ and $|Z_1| < z_{a/2}$, where $a$ is level of significance. For $a = 0.05$, $z_{a/2} \sim 2$.

*Power of CMT1*

Power $= P(Z_2 > 2, Z_1 < 2; \rho, \mu_2 \neq 0, \mu_1 = 0)$

$$= \int_{z_1=-2}^{2} \left[ \int_{z_2>2} f(z_1, z_2) dz_2 + \int_{z_2<-2} f(z_1, z_2) dz_2 \right] dz_1$$

Since conditional distribution of $Z_2 | Z_1 = z_1 \sim N(\mu_2 + \rho(z_1 - \mu_1), 1 - \rho^2)$

$P(|Z_1| < 2, |Z_2| > 2)$

$$= \int_{z_1=-2}^{2} [P(Z_2 > 2 | Z_1 = z_1) + P(Z_2 < -2 | Z_1 = z_1)] f_{Z_1}(z_1) dz_1$$

where $f_{Z_1}$ is pdf of $N(\mu_2 + \rho(z_1 - 0), 1 - \rho^2)$. So,

Power $= \int_{z_1=-2}^{2} \left[ 1 - \Phi\left(\frac{2-\mu_2-\rho z_1}{\sqrt{1-\rho^2}}\right) + \Phi\left(\frac{-2-\mu_2-\rho z_1}{\sqrt{1-\rho^2}}\right) \right] \phi(z_1) dz_1$ -- (1)

where $\Phi(z_1)$ and $\phi(z_1)$ are the standard normal CDF and PDF.

Note that for $\rho = 0$, $P(|Z_1| < 2, |Z_2| > 2; \mu_2) = P(|Z_1| < 2) \cdot P(|Z_2| > 2; \mu_2) = [2\Phi(2) - 1] \cdot [1 - \Phi(2 - \mu_2) + \Phi(-2 - \mu_2)] \approx 0.9545 \cdot [1 - \Phi(2 - \mu_2) +$

$\Phi(-2-\mu_2)]$.

Fig 2. shows the probability of rejecting $H_0$: non-CM vs. $u_2$ for various settings of $(Z_1, Z_2) \sim BVN(\mu_1, \mu_2, 1, 1, \rho)$. The probabilities were calculated via numerical integration using (1). In particular, the line with triangular legend with $\mu_1 = 0$ represent the power function of identifying CM. It coincides with heuristics that the power increases as $\mu_2$ increases. Values of $\rho$ does not seems to affect this probability when $\mu_1 = 0$ and lines for $\rho$ at 0, 0.3, 0.5 all overlapped together. At $\mu_2 = 2$, the power of detecting CM is only 0.4773 (=0.9545*0.5), while at $\mu_2 > 3$ and $> 4$, the powers $> 0.8$ and $> 0.93$, respectively.

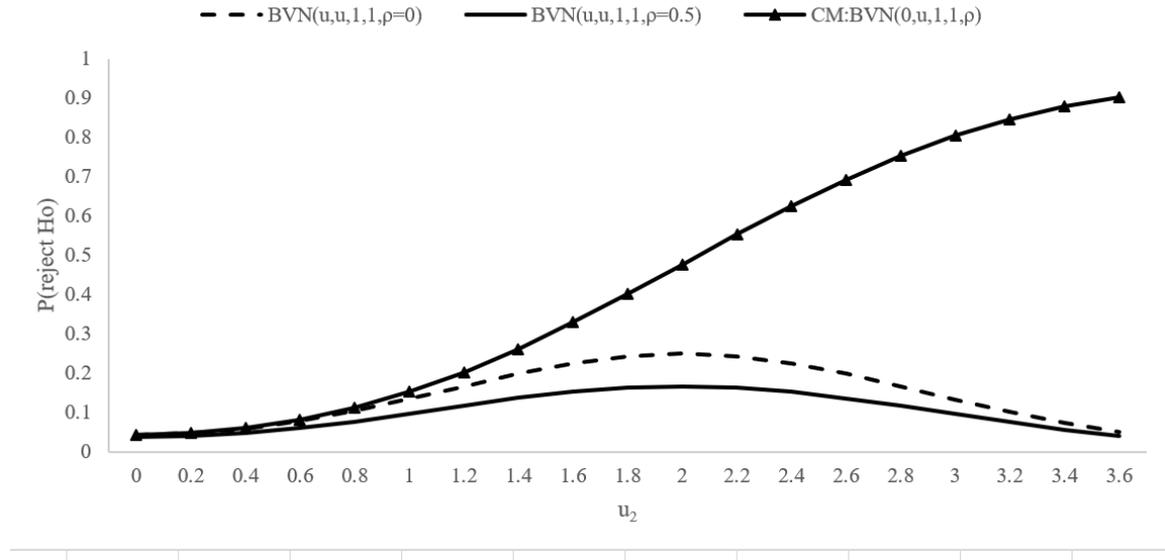

Fig 2. Probability of rejecting $H_0$: non-CM vs. $u_2$ for various $(Z_1, Z_2) \sim BVN(\mu_1, \mu_2, 1, 1, \rho)$.

*Type I error rate of CMT1*

The following theorem shows that for non-CM special case with $E[Z_{IE}] = E[Z_{DE}] = u \neq 0$, the Type I error of CMT1, i.e., $P(\text{reject } H_0: \text{non-CM}) = P(|Z_{DE}| < 2 \text{ and } |Z_{IE}| > 2; \rho, u \neq 0)$, has maximum 0.25 at $\rho = 0$ and $u = \pm 2$.

**Theorem.** For $(Z_1, Z_2) \sim \text{BVN}(u, u, 1, 1, \rho)$, $P(|Z_1| < 2, |Z_2| > 2; \rho, u)$ is maximized at $\rho = 0$ and $u = 2$ with value 0.25. (Proof in Appendix)

In Fig 2., the dash and solid lines at lower area of figure represent the two special cases of non complete mediation with $(Z_1, Z_2) \sim \text{BVN}(u, u, 1, 1, \rho)$ at $\rho = 0$ and 0.5. From the theorem above, the type I error rate by CMT1 can reach 25%. The purpose of adding additional criterion with threshold for $\frac{|Z_1|}{|Z_2|}$ or $\frac{|Z_1|}{|Z_1|+|Z_2|}$ is to reduce the type I error while with reducing as least power as possible.

**Proportion of mediation**

To remedy the insufficiency of criterion (ii), a third criterion utilizing proportion mediated (PM),

$$r = \frac{\text{Indirect effect}}{\text{Total effect}} = \frac{\alpha\beta}{\alpha\beta + \tau'} = 1$$

is often used to align conceptually with the complete mediation. Kenny & Judd (2014) proposed a heuristic rule: if $r > \delta$ for a chosen constant $\delta$ in addition to (i) and (ii), complete mediation may be inferred. They recommended $\delta$=0.8 based on experiences, although this rule has not yet been validated through simulation studies.

Baron and Kenny (1986) classified mediation into two categories, *Complementary mediation,* where both mediated effect (αβ) and direct effect ($\tau'$) exist and point in the same direction, and *Competitive mediation,* where both effects exist but point in the opposite direction. In cases of competitive mediation, it is possible for the total effect $|\hat{\alpha} \cdot \hat{\beta} + \hat{\tau}'|$ to be smaller than the indirect effect $|\hat{\alpha} \cdot \hat{\beta}|$, or even approach zero. This can result in a PM ($r$) greater than one. In such scenarios, a counterintuitive outcome may occur where the total effect is statistically non-significant, while the indirect effect is significant.

To address this issue, Zhao et al. (2010) argued that testing the total effect is unnecessary for establishing mediation. To avoid the inconsistencies caused by competitive mediation, we adopted the absolute proportion of mediation (APM) in this study, defined as:

$$\hat{r}_{APM} = \frac{|\hat{\alpha} \cdot \hat{\beta}|}{|\hat{\alpha} \cdot \hat{\beta}| + |\hat{\tau}'|}$$

This formulation ensures that the proportion remains bounded between 0 and 1.

It is noted that both PM and APM give equal weight to the estimated direct ($\hat{\tau}'$) and indirect ($\hat{\alpha} \cdot \hat{\beta}$) effects, which lead to potential drawback. Specifically, in criterion (ii), an insignificant direct effect may arise not because the effect size is small, but because the standard error of the estimate, $se(\hat{\tau}')$, is large, i.e., the direct effect estimate $\hat{\tau}'$ may still be sizable but imprecise. As a result, the estimated mediation proportion $\hat{r}_{APM}$ may underestimated due to the large variation of the direct effect. To address this limitation, we propose using standardized estimates of the effects. This leads to the definition of the Standardized Absolute Proportion of Mediation (SAPM), which serves as an alternative criterion:

$$\hat{r}_{SAPM} = \frac{|\hat{\alpha} \cdot \hat{\beta}/se(\hat{\alpha} \cdot \hat{\beta})|}{|\hat{\alpha} \cdot \hat{\beta}/se(\hat{\alpha} \cdot \hat{\beta})| + |\hat{\tau}'/se(\hat{\tau}')|}.$$

This formulation reflects the relative strength of the indirect and direct effects in standardized units, accounting for their variability. The criteria for complete mediation test (CMT) are constructed as follows:

(i) Significant Indirect Effect (Evidence of Mediation): $|Z_{IE}| = \frac{|\hat{\alpha} \cdot \hat{\beta}|}{se(\hat{\alpha} \cdot \hat{\beta})} > z_{0.025}$,

(ii) Non-significant Direct Effect: $|Z_{DE}| = \frac{|\hat{\tau}'|}{se(\hat{\tau}')} < z_{0.025}$.

(iii) Dominance of Indirect Effect (Proportion-Based Condition): $\hat{r}_{APM} > \delta$, or

(iii') $\hat{r}_{SAPM} > \delta$.

Here δ represents a user-defined cutoff indicating the minimum acceptable proportion of the total effect that must be mediated.

Based on the criteria above, we define three versions of complete mediation (CM) test as follows:

CMT1: CM is established if both (i) and (ii) hold (significance-based criterion);

CMT2($\delta$): CM is established if (i), (ii), (iii) hold (additional APM criterion), and

CMT3($\delta$): CM is established if (i), (ii), (iii') hold (additional SAPM criterion).

**4. Reduction of Type I error rate by SAPM**

In this session we used numerical integration to compare the Type I error rate of CMT1 and CMT3($\delta$). Let $Z_1 = Z_{DE}$, $Z_2 = Z_{IE}$, and assume $(Z_1, Z_2) \sim$ BVN$(u_1, u_2, 1, 1, \rho)$ with pdf $f(z_1, z_2)$. Given $u_1 \neq 0$, P(reject $H_0$: non–CM) is the Type I error rate for CMT1 and CMT3($\delta$) respectively such that:

$$P(|Z_{DE}| < 2 \text{ and } |Z_{IE}| > 2; \rho, u) = \iint_{|z_2|>2, |z_1|<2} f(z_1, z_2) dz_1 dz_2$$

and

$$P\left(|Z_{DE}| < 2, |Z_{IE}| > 2, \frac{|Z_{IE}|}{|Z_{IE}|+|Z_{DE}|} > \delta; \rho, u\right) = \iint_{|z_2|>2, |z_1|<2, \frac{|z_1|}{|z_2|}<r} f(z_1, z_2) dz_1 dz_2$$

Note that the PSAM $= \frac{|Z_{IE}|}{|Z_{IE}|+|Z_{DE}|} > \delta$ is equivalent to $\frac{|Z_{DE}|}{|Z_{IE}|} < r$ when $r = \frac{1-\delta}{\delta}$.

Fig 3. shows the Type I error rates for CMT1 and CMT3($\delta$) with $\delta$=0.8 and 0.67. Both settings $(u_1, u_2) = (1,1)$ and $(u_1, u_2) = (1,2)$ represent non-CM situations. As expected, CMT1 misclassifies the cases as CM with probabilities ranging from 0.1 to 0.25 and 0.15 to 0.25 respectively. In contrast, both CMT3(0.8) and CMT3(0.67) successfully reduce the error rate to below 0.05.

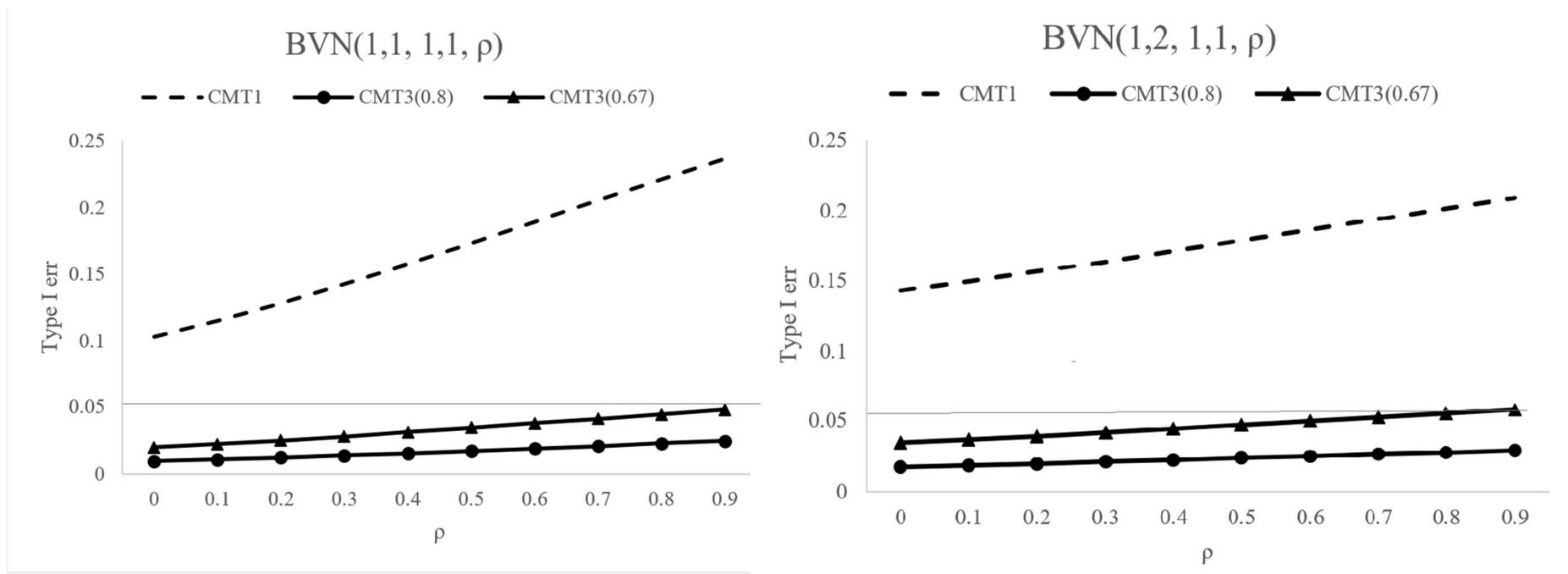

Fig 3. Type I error rate for CMT1 and CMT3($\delta$) under $(Z_1, Z_2) \sim \text{BVN}(u_1, u_2, 1, 1, \rho)$

## 5. Dichotomous mediator and response

When M or Y are dichotomous, the results from causal mediation analysis can differ substantially from those of traditional linear mediation models. In the absence of interaction between exposure and mediator, the natural indirect effect and natural direct effect are defined as $NIE = E[Y_{A=0,M_{A=1}} - Y_{A=0,M_{A=0}}]$ and $NDE = E[Y_{A=1,M_{A=0}} - Y_{A=0,M_{A=0}}]$ respectively. These effects are typically quantified using odds ratios $OR_{NIE}$ and $OR_{NDE}$, and then the total effect is $OR_{TE} = OR_{NDE} \cdot OR_{NIE}$. The proportion mediated on the log-odds scale is $\frac{\log(OR_{NIE})}{\log(OR_{NIE}) + \log(OR_{NDE})}$.

Analogy to the continuous scenario in previous section, we define the APM for dichotomous M and Y as

$$\hat{r}_{APM} = \frac{|\log(OR_{NIE})|}{|\log(OR_{NIE})| + |\log(OR_{NDE})|}$$

to address the possibility of competitive mediation. To incorporate effect sizes and their variability, standardized test statistics are computed as:

$$Z_{NIE} = \frac{\log(OR_{NIE})}{se(\log(OR_{NIE}))} \text{ and } Z_{NDE} = \frac{\log(OR_{NDE})}{se(\log(OR_{NDE}))}$$

and the standardized APM is calculated by

$$\hat{r}_{SAPM} = \frac{|Z_{NIE}|}{|Z_{NIE}| + |Z_{NDE}|}$$

and the criteria can be expressed as

(i) $|Z_{NIE}| > z_{0.025}$
(ii) $|Z_{NDE}| < z_{0.025}$
(iii) $\hat{r}_{APM} > \delta$ or
(iii') $\hat{r}_{SAPM} > \delta$,

and CMT1, CMT2($\delta$) and CMT3($\delta$) are defined accordingly.

## 6. Simulation study

In this section, we conduct a simulation study to evaluate the performance of the proposed complete mediation tests (CMT1, CMT2($\delta$), and CMT3($\delta$)) under various conditions. Specifically, we assess how different choices of the threshold parameter δ affect the power and type I error rates of the tests, and determine which value of δ yields better results.

6.1. Data-Generating Mechanism

For continuous Y, M and A, we adapted the data-generating mechanism used by Fulcher, et al. (2020) as follows:

$$\begin{cases} A = \Upsilon_1 C_1 + \Upsilon_2 C_2 + \Upsilon_3 C_3 + e_1, \\ M = \alpha X + \theta_1 C_1 + \theta_2 C_2 + \theta_3 C_3 + e_2, \\ Y = \tau' A + \beta M + \varphi_1 C_1 + \varphi_2 C_2 + \varphi_3 C_3 + e_3, \end{cases} - (1)$$

where $C_1 \sim \text{Bernoulli}(0.6)$, $C_2|C_1 \sim \text{Bernoulli}\{\text{expit}(1 + 0.5 \cdot C_1)\}$, $C_3 \sim \text{Bernoulli}(0.3)$, $C_4 \sim N(0,1)$ are 4 covariates, $e_1 \sim N(0,1)$, $e_2 \sim N(0,1)$, $e_3 \sim N(0,4)$ are independent white noises, and random coefficients $\Upsilon_1, \Upsilon_2, \Upsilon_3, \theta_1, \theta_2, \theta_3, \varphi_1, \varphi_2, \varphi_3$ follow $N(0,4)$ independently. The indirect effect is determined by $(\alpha, \beta) \in \{(0.3, 0.7), (0.4, 0.6), (0.5, 0.5), (0.8, 0.8)\}$, while direct effect size $\tau'$ was set to be 0 (complete mediation), 0.2 (moderate direct effect), or 0.3 (strong direct effect) respectively. This framework enables a systematic comparison of the CMT criteria under both complete and partial mediation settings, across varying effect sizes and levels of noise.

For dichotomous Y, M and A, the following mechanism was adopted:

$$\begin{cases} X \sim \text{Bernoulli}\{\text{expit}(\Upsilon_1 C_1 + \Upsilon_2 C_2 + \Upsilon_3 C_3)\} \\ M|X, C_1, C_2, C_3 \sim \text{Bernoulli}\{\text{expit}(\alpha X + \theta_1 C_1 + \theta_2 C_2 + \theta_3 C_3)\}, \\ Y|X, M, C_1, C_2, C_3 \sim \text{Bernoulli}\{\text{expit}(\tau' X + \beta M + \varphi_1 C_1 + \varphi_2 C_2 + \varphi_3 C_3)\}. \end{cases} - (2)$$

For each $(\alpha, \beta)$ setting, we generated 2000 datasets, each containing a sample size of 1000, based on model (2). Specifically, 1000 datasets were generated with $\tau' = 0$ (CM), and the other 1000 datasets were generated with $\tau' \neq 0$ (non-CM). For ach dataset we first used LASSO algorithm (Tibshirani, 1994) for covariate selection, conducted mediation analysis, and calculated the corresponding APM and SAPM. We then applied CMT1, CMT2($\delta$) and CMT3($\delta$) to test for complete mediation under hypothesis $H_0$: non-CM vs. $H_1$: CM. CM is identified if $H_0$ was rejected. We varied the threshold $\delta$ in CMT2 and CMT3 to evaluate its influence on performance. Simulation was conducted using SAS/STAT® 15.2, employing the GLMSELECT procedure for covariate selection and the CAUSALMED Procedure for mediation analysis.

6.2 Evaluation Metrics

To evaluate the performance of each criterion, we calculated:

Sensitivity (*sen* = P(Reject H₀ | dataset from CM)): ability to correctly identify CM,

Specificity (*spe* = P(Fail to reject H₀ | dataset from non-CM)): ability to correctly avoid false positives under partial mediation.

For each ($\alpha$, $\beta$, $\tau'$) setting, a receiver operating characteristic (ROC) curves were plotted to illustrate the trade-off between sensitivity and 1 − specificity across a range of δ thresholds. Youden index (sensitivity + specificity − 1) was computed for each δ value. The δ that maximized the Youden Index was considered as the optimal threshold for each criterion.

## 7. Results

7.1 Continuous Mediator and Outcome

Table 1 summarizes the sensitivity and specificity of the criterion CMT1, CMT2($\delta$) and CMT3($\delta$) across different values of $\delta$ in identifying complete mediation under setting of continuous Y and M with parameter configuration ($\alpha$=0.5, $\beta$=0.5, $\tau'$=0 vs 0.2 & 0.3). As expected, CMT1, which applies the least stringent criteria—requiring only conditions (i) and (ii)—achieves the highest sensitivity (0.888), however, at the cost of low specificity, with values of 0.297 and 0.556 for $\tau'$=0.2 and 0.3, respectively.

This indicates a high Type I error rate—misclassifying partial mediation as complete mediation in up to 30–56% of cases.

The inclusion of criterion (iii) for CMT2($\delta$) and criterion (iii′) for CMT3($\delta$) markedly improves specificity by tightening the definition of complete mediation. Although this results in a moderate reduction in sensitivity, the overall performance improves, as indicated by the Youden index. Among them, CMT3($\delta$) consistently outperforms CMT2($\delta$) in terms of the Youden index. Notably, CMT3 with $\delta = 0.75$ or $0.8$ achieves the highest Youden index across both non-zero direct effect settings ($\tau'=0.2$ and $0.3$), suggesting optimal trade-off between sensitivity and specificity.

For other combinations of ($\alpha$, $\beta$, $\tau'$), the performance metrics were visualized using ROC curves, shown in Fig 4 and 5. Across all simulated scenarios, CMT3 with $\delta = 0.75$ or $0.8$ consistently achieves the highest Youden index, indicating robustness discriminative ability in identifying complete mediation.

7.2 Dichotomous Mediator and Outcome

Analogously, Table 2 reports the sensitivity and specificity of CMT1, CMT2($\delta$), and CMT3($\delta$) across a range of $\delta$ values for identifying complete mediation in settings with dichotomous M and Y.

For parameter settings ($\alpha=0.5$, $\beta=0.5$, $\tau'=0$ vs 0.2 & 0.3) and ($\alpha=0.8$, $\beta=0.8$, $\tau'=0$ vs 0.2 & 0.3), as with the continuous case, CMT1 consistently exhibits the highest sensitivity but also the lowest specificity. In contrast, the inclusion of criterion (iii) in CMT2($\delta$) and criterion (iii′) in CMT3($\delta$) substantially increases specificity, although with a modest decrease in sensitivity. Importantly, CMT3($\delta$) again outperforms CMT2($\delta$) in terms of the Youden index, demonstrating superior diagnostic ability. Among the various $\delta$ values tested, CMT3 with $\delta = 0.65$ or $0.7$ achieves the highest

Youden index, indicating that this threshold yields the most effective trade-off between sensitivity and specificity in the dichotomous setting.

Results for other (α, β, τ′) are visualized via ROC curves in Fig 6 and 7, which consistently confirm that CMT3 with δ = 0.65 or 0.7 provides optimal performance across a variety of scenarios.

**8. Testing pleiotropy in real data**

A key step in conducting a Mendelian randomization (MR) study is to identify as many relevant genetic variants (GVs) as possible to serve as instrumental variables for assessing the causal effect of exposure **X** on outcome **Y**. An important criterion for any of the GVs to be valid instruments is the non-pleiotropy condition: a GV should influence the outcome **Y** solely through the exposure **X**—in other words, **X** should complete mediate the effect of the GV on **Y**.

An MR study by Lian et al. (2023) used UK Biobank data to investigate the effect of insomnia on the risk of coronary heart disease (CHD) in menopausal women. First a genome-wide association analysis was conducted in the UK Biobank cohort to identify genetic variants associated with insomnia. Fifty single nucleotide polymorphisms (SNPs) showing the significant associations with insomnia were selected, based on the criteria of an information score >0.8, minor allele frequency >0.01, deletion rate <0.05, and Hardy–Weinberg equilibrium P-value < $10^{-7}$. Among these, 47 SNPs were identified by Lian et al. (2023) as valid instrumental variables after thorough assessment for pleiotropy using available methods, including MR-Egger and Cochran's Q test (Bowden & Holmes, 2019), while the remaining 3 SNPs were deemed invalid and excluded (Listed in Appendix). We then conducted an experiment by applying various CMT to assess the pleiotropy of these 100 SNPs. For each CMT, sensitivity was defined as the proportion of valid SNPs whose effect on CHD was completely mediated by insomnia. Specificity was defined as the proportion of invalid SNPs that were tested as non-CM by the CMT. While effective in detecting global violations of assumptions, omnibus tests are sensitive to outliers or influential points, and may not pinpoint individual SNPs violating the non-pleiotropy assumption. An advantage of CMT is its ability to is its ability to assess pleiotropy at the individual SNP level, offering a more granular analysis.

Table 3 presents the summary of sensitivity of CMT1, CMT2(δ), and CMT3(δ) in detecting CM among 47 valid SNPs. CMT2 has very low sensitivity due to severely under-estimation of the proportion of mediation. Both CMT1 and CMT3 has nearly perfect sensitivity, however, CMT1 falsely detects one among the 3 invalid SNPs as CM.

For illustration the detail of each test, Table 4 presents results of two of the valid SNPs—*rs12094039_T* and *rs113851554_T*—and one of the invalid SNP, *rs112270607_T*, to assess the mediation pathway SNP → Insomnia → CHD. In these analyses, both insomnia and CHD were binary, and SNPs were ternary variables (0, 1, 2). Causal mediation analyses were conducted using SAS/STAT® 15.2 CAUSALMED Procedure. The results illustrate that CMT3 not only supports established findings from omnibus tests but also enables identification of pleiotropic variants that may bias MR conclusions if left unaccounted for, whereas CMT2 exhibits higher misclassify rate. Thus, CMT3 serves as a complementary tool for strengthening the validity of MR studies through individual-level pleiotropy detection.

## 9. Discussion

While testing the hypothesis $H_0$: *non-complete mediation vs.* $H_1$: *complete mediation*, relying solely on the two naïve criteria—(i) a statistically significant natural indirect effect (NIE), and (ii) a non-significant natural direct effect (NDE)—may lead to severely inflated Type I error rates. To mitigate this risk, a third criterion is essential: (iii) evidence of dominance of indirect effect over the direct effect. Kenny & Judd (2014) proposed using proportion-based criterion, such as NIE/(NIE + NDE) > δ = 0.8. However, this proportion can be misleading in scenarios of competitive mediation, where indirect and direct effects act in opposite directions. Furthermore, the threshold of 0.8 was empirically motivated for settings of continuous mediator and outcome, and lacks formal validation through simulation studies.

In this study we propose CMT3, a procedure incorporates Standardized Absolute Proportion of Mediation ($\hat{r}_{SAPM}$), and compare its performance against CMT2, which uses the unstandardized Absolute Proportion of Mediation ($\hat{r}_{APM}$) across a range of δ thresholds. Our simulations demonstrate that $\hat{r}_{SAPM}$ outperforms $\hat{r}_{APM}$ in terms of the Youden index (sensitivity+specificity-1), thereby reducing the likelihood of false positives in declaring complete mediation. This improvement arises because a non-significant NDE may reflect large standard errors rather than a truly negligible direct effect.

We also find that the optimal threshold δ depends on the nature of the mediator and outcome. Specifically, for continuous Y and M, δ=0.75 or 0.8 yields the highest Youden index, whereas for binary Y and M, δ=0.65 or 0.7 is optimal. This is because of PM and APM tend to underestimate the extend of mediation when Y and M are binary. Applying the same threshold used for continuous variables in binary settings can thus lead to substantially reduced detection power. Selecting appropriately calibrated thresholds is therefore essential to maintain sensitivity without inflating false positives.

Application of CMT3 to real data from a Mendelian randomization (MR) study further illustrates its utility. Using SNPs as instrumental variables to assess the causal effect of insomnia on coronary heart disease (CHD), CMT3 with $\hat{r}_{SAPM} > \delta$ (at the optimal threshold of 0.65–0.7) not only supports established findings from omnibus methods but also identifies pleiotropic variants that could bias MR conclusions if unaccounted for. Thus, CMT3 serves as a complementary tool for strengthening the validity of MR studies through individual-level pleiotropy detection.

There are alternative statistics to assess the dominance the dominance of indirect effect beside PM, APM and SAPM, for example, proportion of squared standardized effect: $\hat{r}_{PSSE} = \frac{Z_{NIE}^2}{Z_{NIE}^2 + Z_{NDE}^2}$. This measure follows a non-central Beta(0.5, 0.5, $\lambda$) distribution when $Z_{NIE} \sim N(\lambda, 1)$ and $Z_{NDE} \sim N(0,1)$ are independent. However, in practice, the independence assumption is problematic and the non-centrality $\lambda$ is unknown. Our simulation indicate that this measure can be highly right-skewed and thus highly sensitive to the choice of threshold, with performance no better than $\hat{r}_{SAPM}$.

A limitation of this study is that we only examined scenarios where both the mediator and the outcome are either both continuous or both binary. Further research is needed to address cases involving mixed variable types.

**Ethical statement**

The original study (Lian, et al., 2023) was conducted using UK biobank Research under application number 46789. The research has been approved by the Institute Review Board of Biomedical Sciences, Academia Sinica, Taipei, Taiwan. In the present work, we use a delinked subsample solely for methodological illustration.

Table 1. Sensitivity and specificity of CMT1, CMT2($\delta$) and CMT3($\delta$) under setting of continuous Y and M with ($\alpha$=0.5, $\beta$=0.5, $\tau'$=0 vs 0.2 & 0.3).

| | $\alpha = 0.5, \beta = 0.5$ | | | | |
|---|---|---|---|---|---|
| $\tau'$ | 0 | 0.2 | | 0.3 | |
| | sen | 1-spe | Youden | 1-spe | Youden |
| CMT1 | 0.888 | 0.703 | 0.185 | 0.444 | 0.444 |
| CMT2(0.85) | 0.243 | 0.124 | 0.119 | 0.033 | 0.21 |
| CMT2(0.8) | 0.338 | 0.17 | 0.168 | 0.058 | 0.28 |
| CMT2(0.75) | 0.425 | 0.219 | 0.206 | 0.084 | 0.341 |
| CMT2(0.7) | 0.534 | 0.284 | 0.25 | 0.121 | 0.413 |
| CMT2(0.65) | 0.616 | 0.348 | 0.268 | 0.161 | 0.455 |
| CMT2(0.6) | 0.701 | 0.432 | 0.269 | 0.226 | 0.475 |
| CMT2(0.55) | 0.78 | 0.521 | 0.259 | 0.295 | 0.485 |
| CMT2(0.5) | 0.836 | 0.602 | 0.234 | 0.369 | 0.467 |
| CMT3(0.85) | 0.482 | 0.253 | 0.229 | 0.104 | 0.378 |
| CMT3(0.8) | 0.62 | 0.345 | **0.275\*** | 0.155 | 0.465 |
| CMT3(0.75) | 0.738 | 0.469 | 0.269 | 0.246 | **0.492\*** |
| CMT3(0.7) | 0.816 | 0.569 | 0.247 | 0.346 | 0.47 |
| CMT3(0.65) | 0.855 | 0.65 | 0.205 | 0.408 | 0.447 |
| CMT3(0.6) | 0.878 | 0.684 | 0.194 | 0.437 | 0.441 |
| CMT3(0.55) | 0.885 | 0.702 | 0.183 | 0.444 | 0.441 |
| CMT3(0.5) | 0.888 | 0.703 | 0.185 | 0.444 | 0.444 |

*Maximal Youden of column 4 &5 occurred at CMT3(0.8) and CMT3(0.75).

Table 2. Sensitivity and specificity of CMT1, CMT2($\delta$) and CMT3($\delta$) under setting of dichotomous Y and M with ($\alpha$=0.5, $\beta$=0.5, $\tau'$=0 vs 0.2 & 0.3).

| | $\alpha = 0.5, \beta = 0.5$ | | | | |
|---|---|---|---|---|---|
| $\tau'$ | 0 | 0.2 | | 0.3 | |
| | sen | 1-spe | Youden | 1-spe | Youden |
| CMT1 | 0.375 | 0.3 | 0.075 | 0.23 | 0.145 |
| CMT2(0.8) | 0.223 | 0.134 | 0.089 | 0.087 | 0.136 |
| CMT2(0.75) | 0.227 | 0.139 | 0.088 | 0.095 | 0.132 |
| CMT2(0.7) | 0.234 | 0.143 | 0.091 | 0.097 | 0.137 |
| CMT2(0.65) | 0.242 | 0.149 | 0.093 | 0.104 | 0.138 |
| CMT2(0.6) | 0.246 | 0.153 | 0.093 | 0.114 | 0.132 |
| CMT3(0.8) | 0.178 | 0.104 | 0.074 | 0.065 | 0.113 |
| CMT3(0.75) | 0.231 | 0.134 | 0.097 | 0.091 | 0.14 |
| CMT3(0.7) | 0.282 | 0.165 | **0.117*** | 0.119 | 0.163 |
| CMT3(0.65) | 0.326 | 0.21 | 0.116 | 0.151 | **0.175*** |
| CMT3(0.6) | 0.352 | 0.252 | 0.1 | 0.179 | 0.173 |
| | $\alpha = 0.8, \beta = 0.8$ | | | | |
| $\tau'$ | 0 | 0.2 | | 0.3 | |
| | sen | 1-spe | Youden | 1-spe | Youden |
| CMT1 | 0.895 | 0.765 | 0.13 | 0.576 | 0.319 |
| CMT2(0.8) | 0.718 | 0.555 | 0.163 | 0.365 | 0.353 |
| CMT2(0.75) | 0.727 | 0.571 | 0.156 | 0.38 | 0.347 |
| CMT2(0.7) | 0.748 | 0.589 | 0.159 | 0.397 | 0.351 |
| CMT2(0.65) | 0.771 | 0.607 | 0.164 | 0.416 | 0.355 |
| CMT2(0.6) | 0.79 | 0.62 | 0.17 | 0.431 | 0.359 |
| CMT3(0.8) | 0.531 | 0.375 | 0.156 | 0.199 | 0.332 |
| CMT3(0.75) | 0.67 | 0.487 | 0.183 | 0.295 | 0.375 |
| CMT3(0.7) | 0.771 | 0.577 | **0.194*** | 0.377 | **0.394*** |
| CMT3(0.65) | 0.852 | 0.674 | 0.178 | 0.483 | 0.369 |
| CMT3(0.6) | 0.884 | 0.736 | 0.148 | 0.551 | 0.333 |

*Maximal Youden of column 4 &5 occurred at CMT3(0.7) or CMT3(0.65).

Table 3 Sensitivity of each CMTs in detecting complete mediation (CM) among 47 valid SNPs.

| True result | Test $\delta$ | CMT1 | CMT2 | | CMT3 | |
|---|---|---|---|---|---|---|
| | | | 0.65 | 0.7 | 0.65 | 0.7 |
| 47 valid | CM | 47 | 7 | 7 | 47 | 46 |
| | Non-CM | 0 | 40 | 40 | 0 | 1 |
| | Sensitivity* | 100% | 14.89% | 14.87% | 100% | 97.87% |
| 3 invalid | CM | 1 | 0 | 0 | 0 | 0 |
| | Non-CM | 2 | 3 | 3 | 3 | 3 |

* Sensitivity: proportion of the valid SNPs for which each CMT successfully detects as complete mediation (i.e., non-pleiotropy).

Table 4. Complete mediation test results for the effect of insomnia on the causal relation between SNPs and CAD, across 5 different criteria.

| | ln(OR) | se(ln(OR)) | Z | p-value | CMT1 | CMT2(0.65) | CMT2(0.7) | CMT2(0.65) | CMT2(0.7) |
|---|---|---|---|---|---|---|---|---|---|
| rs12094039_T (valid) | | | | | | | | | |
| (i) NIE | -0.0109 | 0.0025 | -4.36 | <.0001[*1] | ○ | ○ | ○ | ○ | ○ |
| (ii) NDE | -0.005 | 0.0844 | -0.0592 | 0.9528 | ○ | ○ | ○ | ○ | ○ |
| (iii)APM(%) | 68.00% | - | - | - | | ○ | X | - | - |
| (iii')SAPM(%) | 99.00% | - | - | - | - | - | - | ○ | ○ |
| CM or not | - | - | - | - | Yes | Yes | No | Yes | Yes |
| rs1393823_C (valid) | | | | | | | | | |
| (i) NIE | 0.0026 | 0.0006 | 4.3333 | <.0001 | ○ | ○ | ○ | ○ | ○ |
| (ii) NDE | -0.0034 | 0.0226 | -0.1504 | 0.8814 | ○ | ○ | ○ | ○ | ○ |
| (iii)APM(%) | 43.00% | - | - | - | - | X | X | - | - |
| (iii')SAPM(%) | 97.00% | - | - | - | - | - | - | ○ | ○ |
| CM or not | - | - | - | - | Yes | No | No | Yes | Yes |
| rs112270607_T (invalid) | | | | | | | | | |
| (i) NIE | -0.0062 | 0.0016 | -3.98 | <.0001 | ○ | ○ | ○ | ○ | ○ |
| (ii) NDE | 0.1189 | 0.0534 | 2.099 | 0.036[*1] | ○ | ○ | ○ | ○ | ○ |
| (iii)APM(%) | 5.00% | - | - | - | - | X | X | - | - |
| (iii')SAPM(%) | 65.00% | - | - | - | - | - | - | ○ | X |
| CM or not | - | - | - | - | Yes | No | No | No | No |

*1 For multiple comparisons involving 100 SNPs, the Bonferroni correction can be applied by using a significance level at 0.05/50=0.001.

*2 Marginally rejected due to SAPM not exceeding 65%.

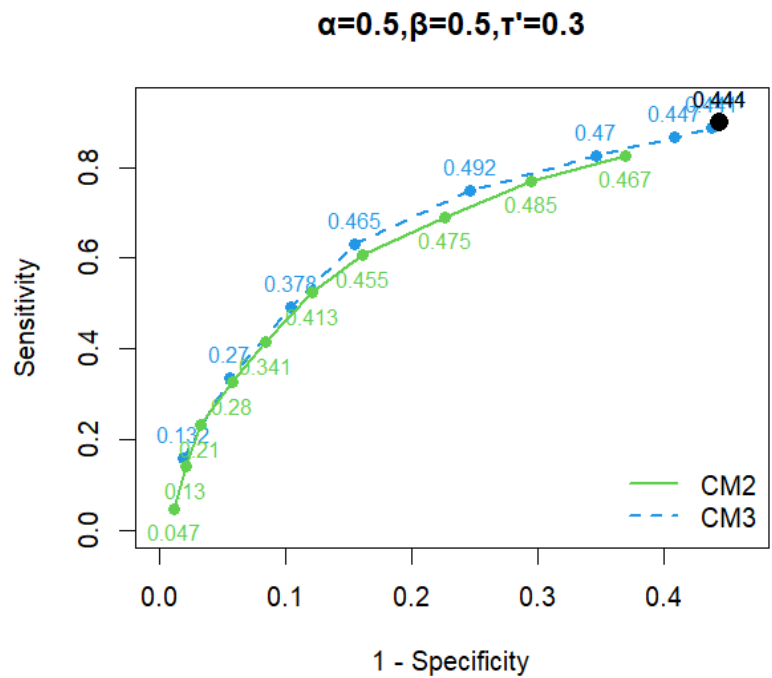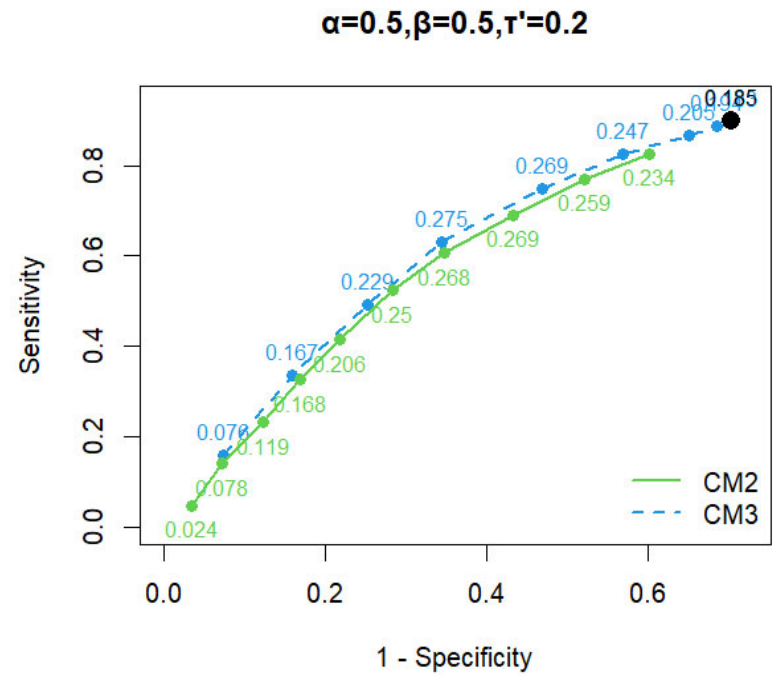

Fig 4. ROC curves and Youden index for continuous Y and M. (1.1) for (α=0.5, β=0.5, τ′=0 & 0.3) and (1.2) for (α=0.5, β=0.5, τ′=0 & 0.2)

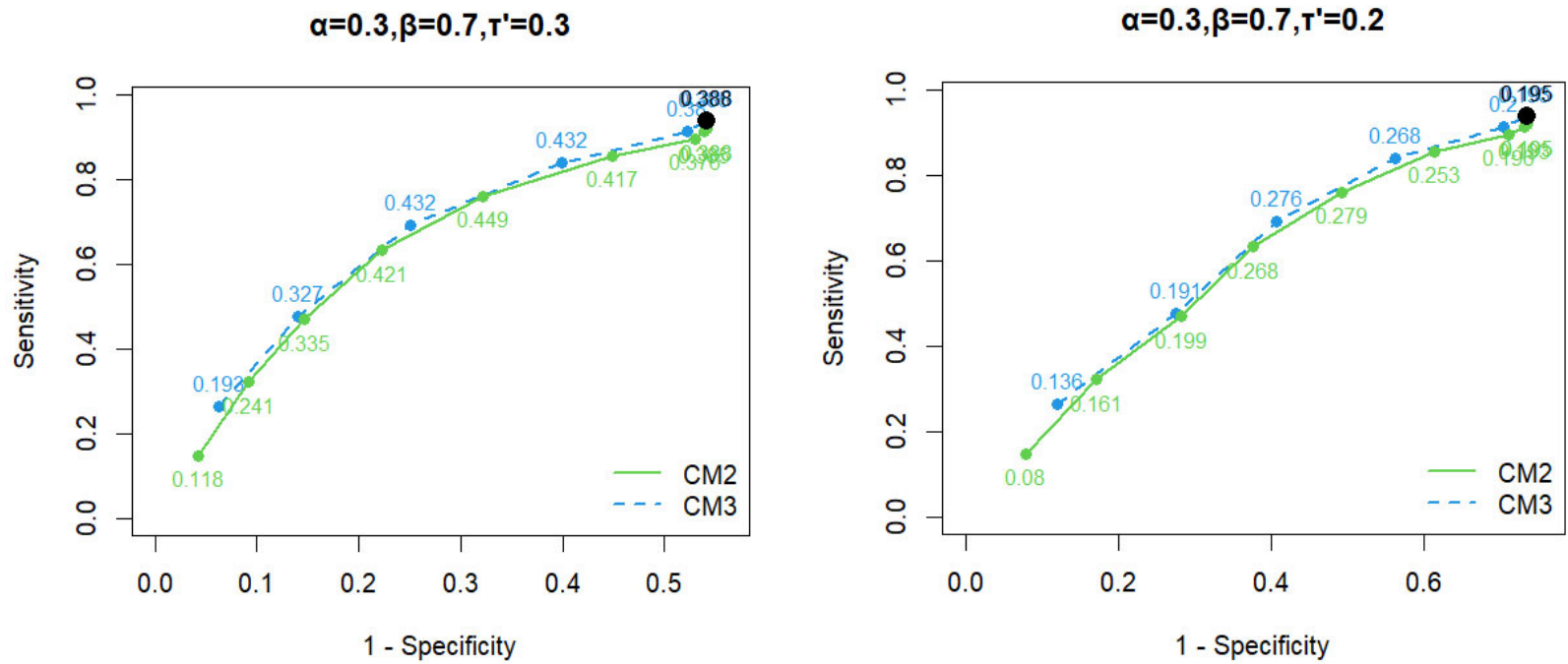

Fig 5. ROC curves and Youden index for continuous Y and M. (2.1) for (α=0.3, β=0.7, τ′=0 & 0.3) and (1.2) for (α=0.3, β=0.7, τ′=0 & 0.2)

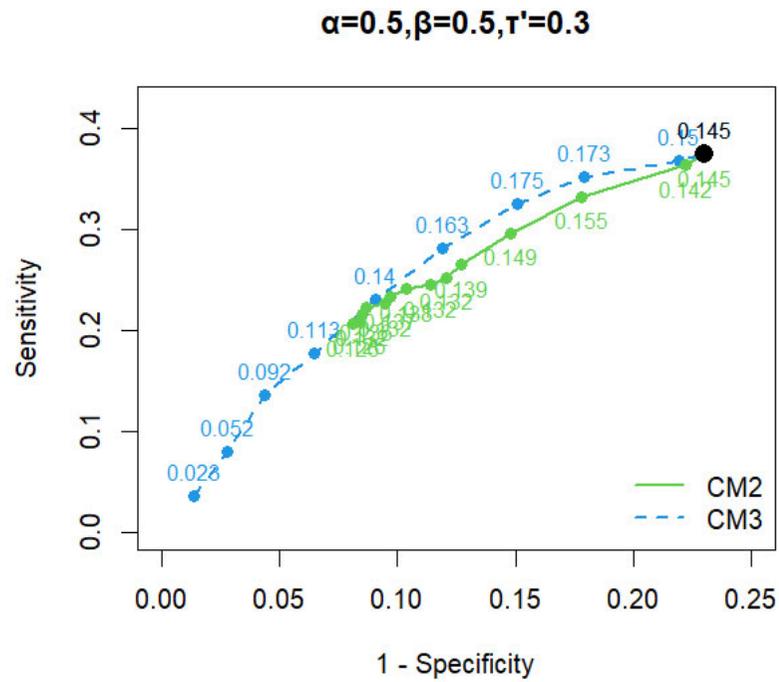 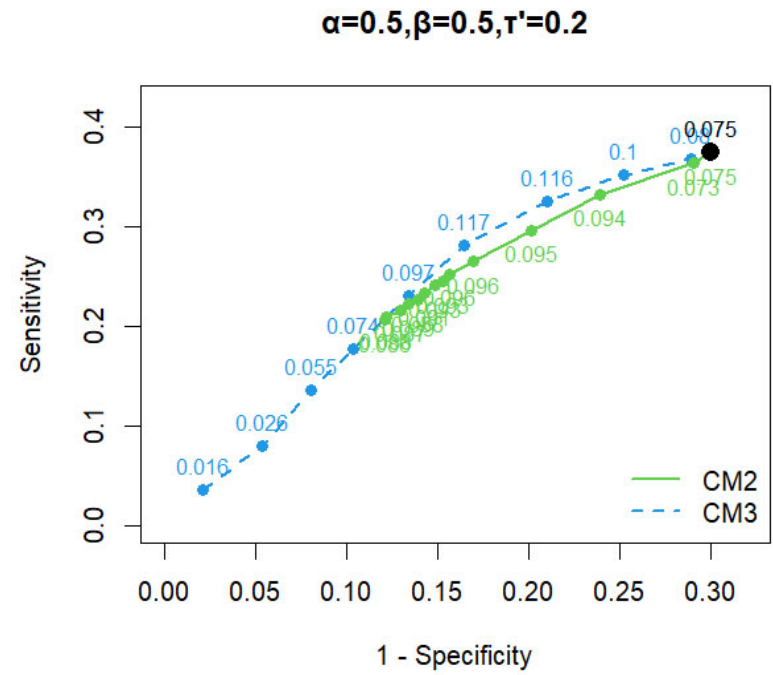

Fig 6. ROC curves and Youden index for dichotomous Y and M. (3.1) for (α=0.5, β=0.5, τ′=0 & 0.3) and (3.2) for (α=0.5, β=0.5, τ′=0 & 0.2)

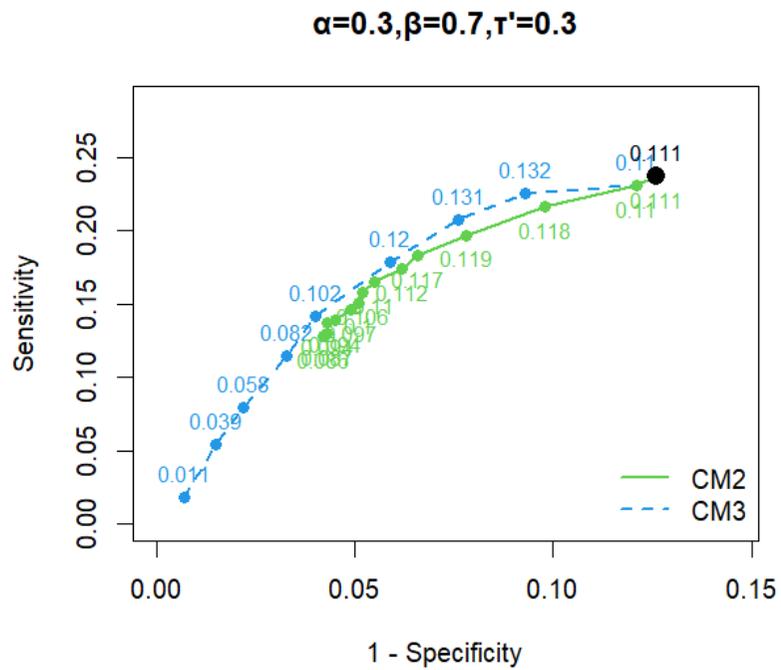 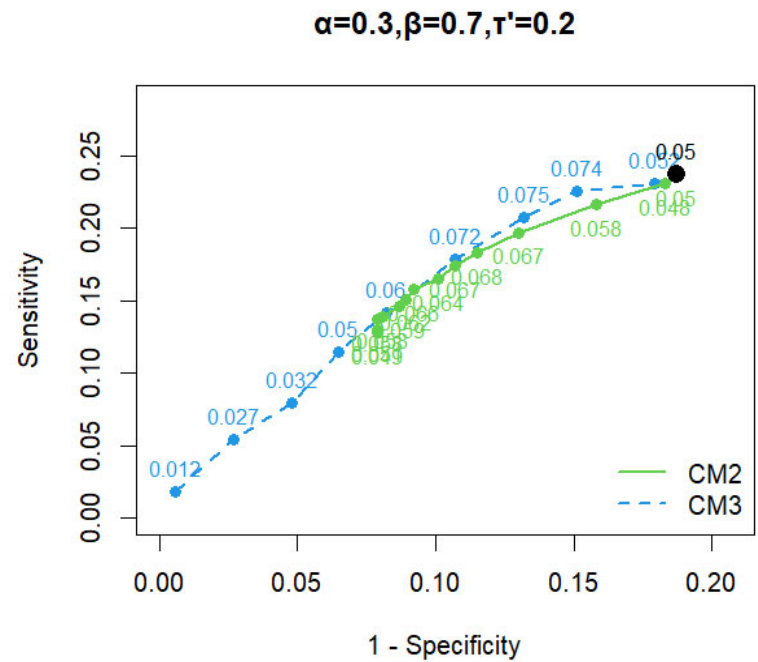

Fig 7. ROC curves and Youden index for continuous Y and M. (4.1) for (α=0.3, β=0.7, τ′=0 & 0.3) and (3.2) for (α=0.3, β=0.7, τ′=0 & 0.2)

APPENDIX 1. Proof of Theorem.

**Theorem.** For $(Z_1, Z_2) \sim \text{BVN}(u, u, 1, 1, \rho)$, $P(|Z_1| < 2, |Z_2| > 2; \rho, u)$ is maximized at $\rho = 0$ and $u = 2$ with value 0.25. (Proof in Appendix)

Proof:

When $\rho = 0$, $P(|Z_1| < 2, |Z_2| > 2; \rho = 0, u) = P(|Z_1| < 2; u) \cdot P(|Z_2| > 2; u)$

Let's call $p(u) = P(|Z| < 2; \mu = u, \sigma = 1)$, $q(u) = P(|Z| > 2; \mu = u, \sigma = 1) = 1 - p(u)$ :

$$p(u) = \Phi(2 - u) - \Phi(-2 - u)$$
$$q(u) = 1 - [\Phi(2 - u) - \Phi(-2 - u)] = \Phi(-2 - u) + 1 - \Phi(2 - u) = 1 - p(u)$$

The product $p(u)q(u) = p(u)(1 - p(u))$ is maximized when $p(u) = 0.5$.

Set $p(u) = \Phi(2 - u) - \Phi(-2 - u) = 0.5$

But since $\Phi(-2 - u) = 1 - \Phi(2 + u)$, the equation becomes

$$\Phi(2 - u) + \Phi(2 + u) = 1.5$$

The function $\Phi(2 - u) + \Phi(2 + u)$ reach 1.5 only when $u = \pm 2$.

For any $\rho \neq 0$, the probability of event $A = \{|Z_1| < 2, |Z_2| > 2; \rho\}$ is the integral of bivariate normal distribution over the shaded regions in Appendix Figure. By *Gaussian Brascamp-Lieb inequality*, the product measure (independence) maximizes the rectangle region among all joint distributions with those marginals, i.e., $P(A; \rho) \leq P(A; 0)$, because any dependency will puts more probability on the corners or on the diagonal (red lines)—reducing the mass inside the shadows rectangle regions.

Therefore we conclude: $\max_{u,\rho} P(|Z_1| < 2, |Z_2| > 2; \rho, u) = 0.25$ at $u = 2, \rho = 0$.

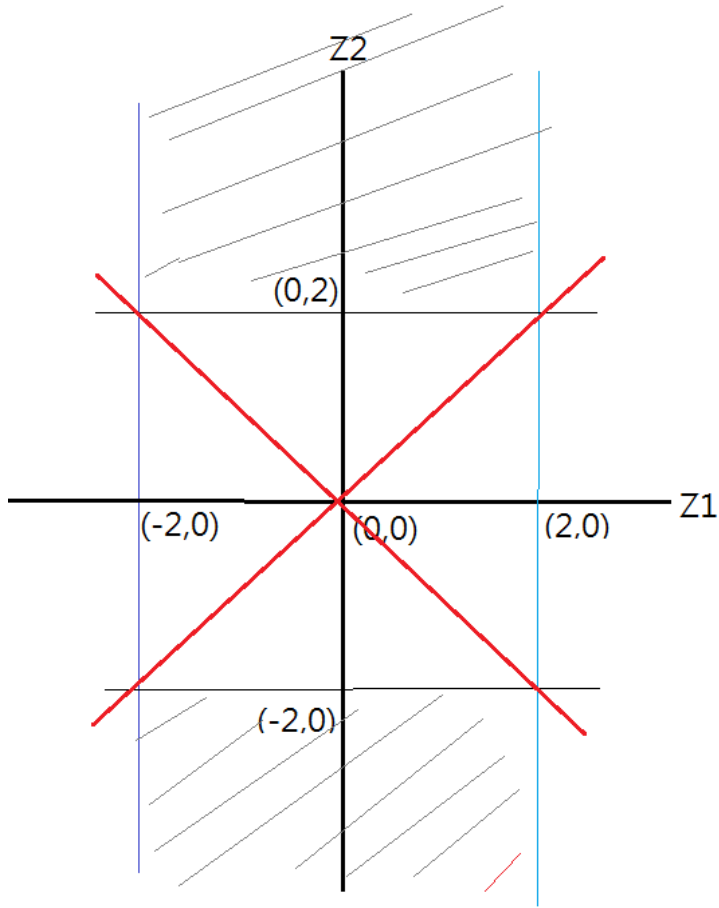

Appendix Figure. Integral of the Bivariate Normal Distribution over Shaded Regions.

Appendix 2. 50 insomnia-associated SNPs as candidates of instrumental variables.

Valid (47)

| | | | |
|---|---|---|---|
| rs12094039_T | rs113851554_T | rs1393823_C | rs111662182_A |
| rs62167787_A | rs409721_T | rs62228444_T | rs1542210_A |
| rs12632133_A | rs35315310_T | rs148058091_C | rs73223324_A |
| rs6855246_G | rs75544266_T | rs34083104_T | rs4569906_C |
| rs115512500_A | rs10046237_G | rs6938026_G | rs13203948_C |
| rs11764779_G | rs940434_C | rs145881501_C | rs6601444_T |
| rs13260185_C | rs35558240_T | rs10983486_C | rs139386514_T |
| rs116894776_T | rs4300383_G | rs641325_T | rs4073582_A |
| rs35929108_A | rs2884799_A | rs4943439_T | rs2759700_T |
| rs10132715_G | rs272801_C | rs146851283_T | rs561411072_T |
| rs145629301_A | rs138539608_G | rs12599233_T | rs7215357_A |
| rs925607_G | rs79311931_G | rs1153416_T | |

Invalid (3)

| | | |
|---|---|---|
| rs114443520_C | rs80259440_A | rs112270607_T |